\documentclass[12pt]{article}
\usepackage[english]{babel}
\usepackage{amsfonts}
 \usepackage[cp1251]{inputenc}
 \usepackage{eufrak}
\begin{document}
 \sloppy

\title{On Some Nonlinear Integral Equation \\
 in the (Super)String Theory}
\author{D.V. Prokhorenko \footnote{Steklov Mathematical Institute, Russian
 Academy of Sciences Gubkin St.8, volovich@mi.ras.ru}}
\maketitle
\begin{abstract}
In this work some nonlinear integral equation is studied. This
equation has arisen in the (super)string field theory and
cosmology. In this work it is proved that some boundary problem
for this equation has a solution.
\end{abstract}
\newpage
\section{Introduction}
Some resent research in (super)string theory \cite{1} leads to
consideration the following pseudo-differential equation:
\begin{eqnarray}
(-q^2\partial^2+1)e^{\partial^2}\Phi(x)=\Phi^3(x)
\label{star},\nonumber\\
q,\;x \in \mathbb{R},\; q>0.
\end{eqnarray}
This equation of motion is obtained in \cite{AJK}. It has been
studied numerically in \cite{yar} where in particular a bound for
the critical value of \(q\) was obtained. A generalization of this
equation to non-flat background has been proposed as a model of
cosmological dark energy \cite{IA}. In \cite{3} it was studied the
energy conservation low for this equation. Let us note that a
quadratic term in the right hand side appears for open bosonic
string and this equation has been studied in \cite{MZ, VSV,
Trento-group}.

In \cite{2} it was studied the equation (\ref{star}) for the
partial case when \(q=0\). There it was shown that this equation
has a solution satisfying the following boundary conditions:
\begin{eqnarray}
\lim_{x\rightarrow +\infty} \Phi(x)=1, \nonumber\\
\lim_{x\rightarrow -\infty} \Phi(x)=-1. \label{bond}
\end{eqnarray}
In this paper we prove that for enough small \(q>0\) the boundary
problem (\ref{bond}) for equation (\ref{star}) has a solution. To
prove this theorem the Leray---Schauder---Tikhonov stable point
theorem is used.
\section{ The main theorem}
We consider the following equation
\begin{eqnarray}
(-q^2\partial^2+1)e^{\partial^2}\Phi(x)=\Phi^3(x)
,\nonumber\\
 q,\;x \in\mathbb{R},\; q\geq0\label{star1}
\end{eqnarray}
in the space \(C^0(\mathbb{R})\) of all bounded real-valued
continuous functions on \(\mathbb{R}\).
 \begin{eqnarray}
 C^0(\mathbb{R})=\{f(x) \in C(\mathbb{R})|\exists
 C>0:\,\forall x \in \mathbb{R}\;|f(x)|<C\},
 \end{eqnarray}
where \(C(\mathbb{R})\) is a space of all real-valued continuous
function on \(\mathbb{R}\). The equation (\ref{star1}) is a formal
form of the following equation:
\begin{eqnarray}
\int \limits_{-\infty}^{+\infty} K_q(x-y)\Phi(y) dy=\Phi(y)^3
\label{star2},
\end{eqnarray}
 where
\begin{eqnarray}
K_q(x-y)=K^0(x-y)+q^2K^1(x-y), \\
K^0(x-y)= \frac{1}{2\sqrt{\pi}}e^{-\frac{(x-y)^2}{4}},\\
K^1(x-y)=\frac{1}{2\sqrt{\pi}}\{\frac{1}{2}-\frac{(x-y)^2}{4}\}e^{-\frac{(x-y)^2}{4}}.
\end{eqnarray}
Let \(T^0\) and \(T^1\) be linear operators on \(C^0(\mathbb{R})\)
defined by the following formulas:
\begin{eqnarray}
(T^0\Phi)(x)=\int \limits_{-\infty}^{+\infty}
K^0(x-y)\Phi(y)dy, \nonumber\\
(T^1\Phi)(x)=\int \limits_{-\infty}^{+\infty}K^1(x-y)\Phi(y)dy,
\end{eqnarray}
and
\begin{eqnarray}
T_q=T^0+q^2T^1.
\end{eqnarray}
 Let \(P^q:C^0(\mathbb{R})\rightarrow C^0(\mathbb{R})\) defined
by the following formula:
\begin{eqnarray}
(P_q\Phi)(x)=(T_q\Phi)^{{1}/{3}}(x).
\end{eqnarray}

\textbf{Theorem.} There exists a real number \(q_0>0\) such that
for all \(0\leq q<q_0\) equation (\ref{star2}) has a bounded
continuous solution \(\Phi(x)\) such that \(\Phi(x)=-\Phi(x)\) and
the following boundary conditions hold:
\begin{eqnarray}
\lim_{x\rightarrow +\infty} \Phi(x)=1, \nonumber\\
\lim_{x\rightarrow -\infty} \Phi(x)=-1. \label{bond1}
\end{eqnarray}

Note that to prove the theorem we use the
Leray---Schauder---Tikhonov stable point theorem.

\textbf{Proof.} We decompose the proof of the theorem into the
sequence of several lemmas.

\textbf{Lemma 1.} There exists a constant \(C^0>0\) such that for
any \(q\in [0,1]\) and for any bounded continuous function
\(\Phi(x)\) satisfying the condition
\begin{eqnarray}
\sup \limits_{x \in \mathbb{R}} | \Phi(x)|\leq C^0
\end{eqnarray}
we have
\begin{eqnarray}
|(P_q\Phi)(x)|\leq C^0\;\forall x \in \mathbb{R}.
\end{eqnarray}

\textbf{Proof.} Let \(\Phi(x) \in C^0(\mathbb{R})\). Let
\(A:=\sup\limits_{x \in \mathbb{R}} | \Phi(x)|\). We have
\begin{eqnarray}
|(T_q\Phi)(x)|\leq A \int \limits_{-\infty}^{+\infty}|K_q(x)|dx.
\end{eqnarray}
Let \(B:=\sup \limits_{q\in [0,1]}\int
\limits_{-\infty}^{+\infty}|K_q(x)|dx\). We get
\begin{eqnarray}
|(P_q\Phi)(x)|\leq B^{1/3}A^{1/3}. \label{in}
\end{eqnarray}
Put \(C^0=B^{1/2}\). It follows from (\ref{in}) that condition
\(\sup \limits_{x \in \mathbb{R}} | \Phi(x)|\leq C^0\) implies
that \(|(P_q\Phi)(x)|\leq C^0\;\forall x \in \mathbb{R}.\) The
lemma is proved.

\textbf{Lemma 2.} There exists a constant \(C^1\) such that for
any \(q\in [0,1]\) and for any continuous function
\(\Phi(x):|\Phi(x)|<C^0\) \(\forall x\), the following statement
holds
\begin{eqnarray}
\forall x',\;x'' \in
\mathbb{R}\;|(P_q\Phi)(x')-(P_q\Phi)(x'')|<C^1|x'-x''|^{1/3}.
\end{eqnarray}

 \textbf{Proof.} Let us estimate the derivative
\((T_q\Phi)'(x):=\frac{d}{dx}(T_q\Phi)(x)\). We have
\begin{eqnarray}
|(T_q\Phi)'(x)|\leq \int \limits_{-\infty}^{+\infty}|K_q'(x-y)|
|\Phi(y)|dx\leq \nonumber\\
\leq C^0 \int \limits_{-\infty}^{+\infty}|K_q'(x)|\leq C^0
E,\;\forall q\in [0,1],\;\forall x \in\mathbb{R}, \label{18}
\end{eqnarray}
where we put by definition
\begin{eqnarray}
E:=\sup \limits_{q\in [0,1]} \int \limits_{-\infty}^{+\infty}
|K'_q(x)|dx.
\end{eqnarray}
Inequality (\ref{18}) immediately implies that
\begin{eqnarray}
|(T_q\Phi)(x')-(T_q\Phi)(x'')|\leq C^0E|x'-x''| \label{A}
\end{eqnarray}
for all \(x',x''\in \mathbb{R}\).

Note that the following statement holds. There exists a constant
\(\hat{C}\) such that \(\forall\, y',y''\in\mathbb{R}\)
\begin{eqnarray}
|(y')^{1/3}-(y'')^{1/3}|\leq\hat{C}|y'-y''|^{1/3}. \label{B}
\end{eqnarray}
Inequalities (\ref{A}) and (\ref{B}) implies that
\begin{eqnarray}
|(P_q\Phi)(x')-(P_q\Phi)(x'')|\leq\hat{C}(C^0E)^{1/3}|x'-x''|^{1/3}.
\end{eqnarray}
Let us denote by \(C^1\) the constant
\(C^1:=\hat{C}(C^0E)^{1/3}\). We have
\begin{eqnarray}
|(P_q\Phi)(x')-(P_q\Phi)(x'')|\leq C^1|x'-x''|^{1/3}.
\end{eqnarray}
The lemma is proved.

By definition \(C(\mathbb{R})\) is a space of all continuous
functions on \(\mathbb{R}\). The space \(C(\mathbb{R})\) is a
Frechet space with respect to the following set of seminorms
\begin{eqnarray}
\{p_n(f):=\sup_{x \in [-n,n]}|f(x)|\}.
\end{eqnarray}

\textbf{Lemma 3.} Let
\begin{eqnarray}
K:=\{\Phi \in C(\mathbb{R})| \sup \limits_{x \in
\mathbb{R}}|f(x)|\leq C^0,\nonumber\\
 |\Phi(x')-\Phi(x'')|\leq
C^1|x'-x''|^{1/3}\,\forall x',x'' \in \mathbb{R}\}.
\end{eqnarray}
K is a compact.

 \textbf{Proof.}
 This lemma follows from the Arzella --- Ascolli theorem by using
 the Cantor diagonal method.

 Let \(\Psi(x)\) be a function:
\begin{eqnarray}
\Psi(x)=\frac{1}{\sqrt{\pi}}\int \limits_0^x e^{-y^2} dy
\end{eqnarray}

\textbf{Lemma 4.} There exists a constant \(C^2\) and a positive
number \(q_0\) such that for any function \(\Phi(x) \in K\)
satisfying
\begin{eqnarray}
\Phi(x)=-\Phi(-x)
\end{eqnarray}
and
\begin{eqnarray}
 \Phi(x)\geq C^2 \Psi(x)\; \rm if \mit \; x>0
\end{eqnarray}
we have
\begin{eqnarray}
(P_q\Phi)(x) \geq C^2 \Psi(x)\; \rm if \mit \; x>0.
\end{eqnarray}
for all \(q \in [0,q_0]\)

 \textbf{Proof.} Let us calculate
\((T^0\Psi)(x)\). We have
\begin{eqnarray}
(T^0\Psi)'(x)=\frac{1}{2\sqrt{\pi}}\int
\limits_{-\infty}^{+\infty}
e^{-\frac{(x-y)^2}{4}}\Psi'(y)dy=\nonumber\\
=\frac{1}{2\sqrt{\pi}}\int \limits_{-\infty}^{+\infty}
e^{-\frac{(x-y)^2}{4}}
\frac{1}{\sqrt{\pi}}e^{-y^2}dy=\nonumber\\
=\frac{1}{\sqrt{5\pi}}e^{-\frac{1}{5}x^2}.
\end{eqnarray}
Therefore
\begin{eqnarray}
(T^0\Psi)(x)=\frac{1}{\sqrt{5\pi}}\int \limits_0^x
e^{-\frac{1}{5}y^2}dy
\end{eqnarray}
Note that \((T^0\Psi)'(0)=\frac{1}{\sqrt{5\pi}}>0\). One can
easily prove by using this fact that there exists a constant
\(C^3>0\) such that \((T^0\Psi)(x)>2C^3\Psi(x)\) if \(x>0\).

The following statement holds \cite{2}. For any \(\Phi(x) \in
C^0(\mathbb{R})\) such that \(\Phi(x)=-\Phi(-x)\) the condition
\(\Phi(x)\geq\Psi(x)\) if \(x>0\) implies that
\((T^0\Phi)(x)>(T^0\Psi)(x)\) if \(x>0\).

Therefore \(\forall \Phi(x) \in C^0(\mathbb{R})\) such that
\(\Phi(x)=-\Phi(-x)\) and \(\Phi(x)\geq\Psi(x)\) if \(x>0\) one
gets
\begin{eqnarray}
(T^0\Phi)(x)\geq 2 C^3\Psi(x).
\end{eqnarray}
Let us prove that there exists a constant \(C^4\) such that
\(\forall \Phi(x) \in C^0(\mathbb{R})\) such that
\(\Phi(x)=-\Phi(-x)\) and \(\sup \limits_{x \in
\mathbb{R}}|\Phi(x)|<C^0\) the following estimate holds
\begin{eqnarray}
|(T^1\Phi)(x)|\leq C^4 |\Psi(x)|\;\forall x \in\mathbb{R}.
\end{eqnarray}
It is easy to prove that \(\sup \limits_{x \in
\mathbb{R}}|(T^1\Phi)(x)|<\infty\), \((T^1\Phi)(0)=0\) and \(\sup
\limits_{x \in \mathbb{R}} |(T^1\Phi)'(x)|<\infty\). For example
\begin{eqnarray}
 |(T^1\Phi)'(x)|=|\int \limits_{-\infty}^{+\infty}
 (K^1)'(x-y)\Phi(y)dy|\leq \nonumber\\
\leq C^0\int  \limits_{-\infty}^{+\infty}|(K^1)'(y)|dy,
\end{eqnarray}
and
\begin{eqnarray}
\int  \limits_{-\infty}^{+\infty}|(K^1)'(y)|dy<\infty.
\end{eqnarray}
Put by definition
\begin{eqnarray}
A_1:=\inf \limits_{x \in [0,1]}\Psi'(x),\nonumber\\
B_1:=\inf \limits_{x \in [1,+\infty]}\Psi(x),\nonumber\\
A_2:=\sup \limits_{x \in [0,+\infty]} |(T^1\Phi)'(x)|,\nonumber\\
B_2:=\sup \limits_{x \in [0,+\infty]} |(T^1\Phi)(x)|.
\end{eqnarray}

Let us chose \(C^4\) such that \(C^4 A_1>A_2\) and \(C^4
B_1>B_2\). We find
\begin{eqnarray}
|(T^1\Phi)(x)|\leq C^4|\Psi(x)|,\;\forall x \in \mathbb{R}.
\end{eqnarray}
Let \(A\) be a real positive number and  let \(q_0\) be a positive
number such that
\begin{eqnarray}
C^4q_0^2<C^3A.
\end{eqnarray}
(\(q_0\) depends on A.) We get
\begin{eqnarray}
(T_q\Phi)(x)\geq C^3 A \Psi(x)\; \rm if \mit x>0
\end{eqnarray}
for any function \(\Phi(x)\) such that \( \sup \limits_{x \in
\mathbb{R}}|\Phi(x)|<C^0\), \(\Phi(x)=-\Phi(-x)\) and \(\Phi(x)> A
\Psi(x)\; \rm if \mit x>0\) and for any \(q \in [0,q_0]\).

One can easily proof that there exists a positive constant \(L\)
such that \(\Psi^{1/3}(x)>L\Psi(x)\; \rm if \mit x>0\). So for any
positive constant \(A>0\) and the function \(\Phi\) satisfying:
 \(\sup \limits_{x \in \mathbb{R}}|\Phi(x)|\leq C^0\),
 \(\Phi(x)=-\Phi(-x)\) and \(\Phi(x)> A
\Psi(x)\; \rm if \mit x>0\) one gets
\begin{eqnarray}
|(P_q\Phi)(x)|\geq (C^3)^{1/3} L A^{1/3}\Psi(x)\; \rm if \mit x>0.
\end{eqnarray}
Let \(C^2\) be a positive constant such that
\begin{eqnarray}
(C^3)^{1/3} L (C^2)^{1/3}\geq C^2.
\end{eqnarray}
If \(\Phi(x)\) satisfy the conditions of the lemma then
\begin{eqnarray}
|(P_q\Phi)(x)|\geq C^2 \Psi(x)\; \rm if \mit x>0.
\end{eqnarray}
The lemma is proved.

 \textbf{Remark.} We can chose the constant \(C^2\) such that
 \(\sup \limits_{x \in\mathbb{R}} C^2|\Psi(x)|<1 \). Below it will
 be assumed that \(C^2\) satisfy this condition.

\textbf{Lemma 5.} Let \(K_1\subseteq K\) consisting of all
functions \(\Phi(x)\) such that

a) \(\Phi(x)=-\Phi(-x).\)

b) \(\Phi(x)\geq C^2 \Psi(x)\; \rm if \mit x>0\)

 \(K_1\) is a compact.

\textbf{Proof.} \(K_1\) is a closed subset of the compact \(K\).
Therefore \(K_1\) is a compact. The lemma is proved.

\textbf{Lemma 6.} \(K_1\) is a convex set i.e. if
\(\Phi_1,\;\Phi_2 \in K_1\) and \(\alpha_1\), \(\alpha_2\) be real
numbers such that \(\alpha_1>0,\;\alpha_2>0\),
\(\alpha_1+\alpha_2=1\) then \(\alpha_1\Phi_1+\alpha_2\Phi_2 \in
K_1\).

\textbf{Proof.} It is evidence.

\textbf{Lemma 7.} The equation (\ref{star2}) has a continuous
bounded solution \(\Phi(x) \in K_1\) such that
\(\Phi(x)=-\Phi(-x).\)

\textbf{Proof.} Lemma 1, lemma 2 and lemma 4 implies that
\(P_q(K_1)\subseteq K_1\). Lemmas 5, 6 implies that \(K_1\) is a
convex compact in some locally convex space. Therefore the present
lemma follows from the Leray
--- Schauder --- Tikhonov stable point theorem \cite{4}.

\textbf{Lemma 8.} If \(\Phi(x)\in K_1\) is a solution of equation
(\ref{star2}) then
\begin{eqnarray}
\Phi(x)\rightarrow +1; \rm as \mit\; x\rightarrow +\infty \; \rm
and
\mit \nonumber\\
\Phi(x)\rightarrow -1; \rm as \mit\; x\rightarrow -\infty.
\end{eqnarray}

\textbf{Proof.} Note that the following statement holds.

Let \(D\) be a real number \(0<D<1\). There exists a constant
\(0<C^5(D)<1\) such that
\begin{eqnarray}
|x^{1/3}-1|\leq C^5(D)|x-1|
\end{eqnarray}
for all \(x\geq D\). The proof of this statement is simple and
omitted. Let \(\Phi(x) \in K_1\) be a solution of (\ref{star2}).
Let \(l_0\) be an arbitrary positive number. Let
\(D_1=\frac{1}{2}C^2\Psi(l_0)\) and \(\delta_1=\sup \limits_{x\in
[l_0,+\infty)} |1-\Phi(x)|\). If \(\delta_1=0\) the lemma is
proved. Suppose that \(\delta_1\neq 0\).  Let us estimate the
difference \(|(P_q\Phi)(x)-\Phi(x)|\). We have
\begin{eqnarray}
|(T_q\Phi)(x)-\Phi(x)|=|\int \limits_{-\infty}^{+\infty} K_q(x-y)
(\Phi(y)-1) dy|\leq\nonumber\\
|\int \limits_{-\infty}^{l_0}K_q(x-y) (\Phi(y)-1) dy|+\nonumber\\
+|\int \limits_{l_0}^{+\infty}K_q(x-y) (\Phi(y)-1) dy|\leq\nonumber\\
 \leq 2C^0 \int \limits_{-\infty}^{l_0}|K_q(x-y)|dy
 +\delta_1 A^{(q)}, \label{CC}
\end{eqnarray}
where we put by definition
\begin{eqnarray}
A^{(q)}=\int \limits_{-\infty}^{+\infty}|K_q(x)|dx.
\end{eqnarray}
The first term in the last line of (\ref{CC}) (we will denote it
by \(\chi_{l_0}(x)\)) tends to zero as \(x \rightarrow +\infty\).
Note that \(A^{(q)}\rightarrow 1\) if \(q\rightarrow 0\). One can
prove that we can chose the constant \(C^0\) from lemma 1 such
that \(|C^0-1|<1-2D_1\) if \(0<q<q_0\) and \(q_0\) is enough
small. So \(|(T_q\Phi)(x)|>D_1\) if \(q_0\) is enough small number
\(0<q<q_0\) and \(x\) is enough large positive number. Moreover
\begin{eqnarray}
\chi_{l_0}(x)+\delta_1 A^{(q)}\leq \frac{1}{\sqrt{C^5(D_1)}}
\delta_1
\end{eqnarray}
if \(0\leq q<q_0\), \(q_0\) is enough small number and \(x\) is
enough large positive number. So there exist positive numbers
\(l_1>0\) and \(q_0>0\) such that
\begin{eqnarray}
|(P_q\Phi)(x)-1|\leq \sqrt{C^5} \delta_1\; \rm if \mit
\end{eqnarray}
\(0<q<q_0\) and \(x>l_0+l_1\).

We can proof as previous that there exists a positive number
\(l_2\) such that
\begin{eqnarray}
|(P_q\Phi)(x)-1|\leq (\sqrt{C^5})^2 \delta_1\; \rm if \mit
\end{eqnarray}
\(x>l_0+l_1+l_2\).

In general by induction we can find a sequence of positive numbers
\(l_i,\;i=0,1,2,...\) such that
\begin{eqnarray}
|(P_q\Phi)(x)-1|\leq (\sqrt{C^5})^n \delta_1\; \rm if
\mit\nonumber\\
x>\sum \limits_{j=0}^n l_j.
\end{eqnarray}
Therefore \(\Phi(x)\rightarrow +1\) as \(x\rightarrow +\infty\).
The lemma is proved.

Combining the consequences of  lemmas 7, 8 we finish the proof of
the theorem.

\section{Conqlusion}
In the present paper we study the boundary problem (\ref{bond})
for the equation (\ref{star}). We have proved by using the Leray
--- Scauder --- Tikhonov theorem that this problem has a solution.

\section{Acknowledgements}
 I wold like to thank V.S. Vladimirov, I.V. Volovich, I. Ya. Aref'eva, L. V. Joukovskaya
  for very useful discussions.\newline
 \indent This work was partially supported by the Russian
 Foundation of Basis Reasearch (project 05-01-008884), the grand
 of the president of the Russian Federation (project
 NSh-1542.2003.1) and the program "Modern problems of
 theoretical mathematics" of the mathematical Sciences department
 of the Russian Academy of Sciences.
 

\begin{thebibliography}{99999}
 \bibitem{1} M. B. Green, J. H. Scwartz and E. Witten. \textit{Superstring
 Theory}, Cambrige University Press, 1987.
 \bibitem{AJK} I. Ya. Aref'eva, L. V. Joukovskaya, A. S. Koshelev.
 JHEP, {\bf 0309}, 2003, 012; hep-th/0301137.
 \bibitem{yar} Ya. I. Volovich, J. Phys. {\bf A36}, 2003, 8685;
 math-ph/0301028.
 \bibitem{IA} I. Ya. Aref'eva. astro-ph/0410443.
 \bibitem{3} L. V. Joukovskaya, Ya. I. Volovich. math-ph/0308034.
 \bibitem{MZ} N. Moeller, B. Zwiebach. JHEP, {\bf 0210}, 2002,
 034.
 \bibitem{VSV} V. S. Vladimirov. Izvestiya RAN.
 \bibitem{Trento-group} V. Forini, G. Grignani, G. Nardelli;
 hep-th/0502151.
 \bibitem{2} V. S. Vladimirov, Ya. I. Volovich.
 Theor. Math. Phys. 138, 2004, 297-309.
 \bibitem{4} M. Reed, B. Simon. \textit{Methods of Modern
 Mathematical Physics: Functional Analysis, volum 1}, San Diego,
 2 edition, 1980.
 \end{thebibliography}
\end{document}